\documentstyle[12pt]{article}
\begin{document}
\voffset=-1.0truecm
\hoffset=0.0truecm

\begin{center}

{\Large STRANGE HADRONS FROM THE ALCOR }
\medskip

{\Large REHADRONIZATION MODEL }

\vskip 0.8 truecm

{\large
T.S. Bir\'o, P. L\'evai, J. Zim\'anyi\footnote{ The talk was presented
by J. Zim\'anyi on S'95 Conference, Tucson, Jan. 3-7 1995.
E-mail:jzimanyi@sunserv.kfki.hu}}
\vskip 0.6cm

{\it Research Institute for Particle and Nuclear Physics}

{\it Budapest 114., P.O.Box 49, H-1525 Hungary}

\end{center}
\vskip 0.8cm

\noindent {\small
ABSTRACT: Hadron multiplicities --- especially for strange particles ---
are calcula\-ted in the  framework  of  the   algebraic coalescence
rehadronization  model  (ALCOR),  which counts for
redistribution of quarks  into  hadrons for relativistic
heavy-ion collisions.
The influence of Bjorken flow on the final  hadronic
composition are incorporated in the model.  A  comparison is made
with the CERN SPS NA35  $S+S$ and $p+p$ experiments.
The analysis of these experiments with ALCOR shows a strangeness
enhancement for S+S collisions and a possible formation of a sort of
semi-deconfined state of the matter.
Predictions for Pb+Pb collisions (NA49) are also presented. }

\vspace{1.0cm}

\section*{Introduction}

The strongly dynamical process of rehadronization from a thermal
quark-gluon plasma or from any other type of prehadronic quark
matter into the finally observed hadrons in a high-energy
heavy-ion
collision cannot be described in the framework of perturbative
parton dynamics alone \cite{Wang} , \cite{parton} nor by using rate equations
as in quarko- and hadrochemistry \cite{qchem},\cite{hchem}.
Not only does the strength of the interaction leading to valence quark
confinement into hadrons in the final state of this process rule
out any expansion in terms of the QCD coupling constant
$g^2/4\pi$, but also the fact that there are only color singlet hadrons
renders a description using rate equations derived in the infinite
volume limit to be unreliable. In this case  there would always
be a small fraction of free quarks present after a
finite time has elapsed  unless the hadronization cross
section diverges after a finite time. In that case the
use of rate equations is also unjustified.

The idea of redistributing {\it all} the quarks and antiquarks
present in the quark matter at the latest stage of its evolution
between color singlet final states (hadrons) cures these two
problems at the level of a phenomenological model.
It describes the rehadronization as a sudden process.
This assumption unfortunately cannot be checked in the
framework of a quark number redistribution model. For this
purpose a comparison with microscopic dynamical calculations
will be necessary.
If one then compares the predictions of such a simple redistribution of
different quark flavors with experimental findings, especially
with strange or charm to light hadron ratios, one may then have a
basis for judging whether nontrivial collective phenomena at the
{\it quark level} must be considered in these experiments.

In applying rehadronization models one has to supply as input
the number of light and heavy valence quarks and antiquarks, all of which
will be hadronized. This number can be obtained from theoretical models
(e.g. string model \cite{Werner}, color rope model \cite{rope},
thermodynamical models, etc.)
or from a part of the observed data. In the latter case comparison of the
predictions of the model with further experimental data may be used to
prove or disprove the basic assumptions of that model. Here,
applying the {\bf AL}gebraic {\bf CO}alescence {\bf R}ehadronization
model, ALCOR \ \cite{ALCOR}, we shall follow the latter possibility.

\section*{The ALCOR model}

The starting point of the ALCOR model
is a state in which only those quarks and antiquarks which
form the final hadrons are present in a local thermal equilibrium.
We also assume that they are all present, so their number
already account for some of the gluons fragmented earlier.
This assumption is  easily acceptable for  the heavy quarks  for
which any meeting  with their antiparticle  is improbable due  to
their small specific  density.  In the  case of light  constituent
quarks  (counted  after  gluon  fragmentation) the approximate
entropy conservation during  the rehadronization process  leads to
the conservation of  their numbers.  This  picture was  first
formulated in the framework  of the algebraic recombination  model
\cite{recom}.  In this model the  number of a given type of hadron
produced is
proportional to the product  of the numbers of  their constituting
quarks.   This   picture  was  developed   further  by  the
introduction of gluon fragmentation \cite{gfrag}.
 We shall denote the number of produced $u,d,s$ quark pairs,
just before the hadronization, by $N_{u,pair}$, $N_{d,pair}$ and
$N_{s,pair}$, respectively.
The strangeness production factor, $g_s$ is defined as
$g_s = N_{s,pair} / ( N_{u,pair} + N_{d,pair})$. (The definition of factor
$f_s$ in Ref. \cite{ALCOR}  was different from that of $g_s$.)

In the  ALCOR model  we generalize  this further by introducing the
coalescence factors $ C_M (i,j) $, $ C_B (i,j,k) $ for each hadron type
separately. In order to ensure that all quarks are included  in
hadrons  in  the  final  state  we also introduce normalization factors
$b(i)$ associated with each quark flavor  when counting the  quarks
and antiquarks  in hadrons.
Furthermore, a spin degeneracy
factor, $D^{(h)}=2S_h+1$, is also included. Here we shall consider
the simplest hadron multiplets conserving isospin symmetry. These are
the spin 0 and spin 1 meson octets together with the $\sigma$
and $\omega$ isospin singlet meson states. The spin 1/2 octet
and spin 3/2 decouplet baryons and anti-baryons are also included.
Thus the  numbers of  a given sort of
baryon or antibaryon, consisting  of quarks with flavors
$i,j,k$, are given as
\begin{eqnarray}
N_B^{(h)}(i,j,k)&=& D^{(h)} \, C_B(i,j,k) \, b(i) \, b(j) \, b(k) \, N_Q(i) \,
N_Q(j) \, N_Q(k) , \label{S1} \\
N_{\overline B}^{(h)}(i,j,k)&=&
D^{(h)} \, C_{\overline B}(i,j,k) \,{\overline b}(i) \, {\overline b}(j)
\, {\overline b}(k)
 \, N_{\overline Q}(i)\, N_{\overline Q}(j)\, N_{\overline Q}(k) , \label{S2}
\end{eqnarray}
\noindent  and the  number of a given sort of meson is given as
\begin{equation}
N_M^{(h)}(i,j)  = D^{(h)} \, C_M(i,j)\, b(i)\, {\overline b}(j)\, N_Q(i)\,
N_{\overline Q}(j). \label{S3} \\
\end{equation}
Here $N_Q(i)$ and $N_{\overline Q}(i)$ are the number of quarks
and antiquarks of type $i$ , respectively.

These constraints lead finally to the following system of equations  (one
for each quark and each antiquark flavor)
\noindent
\begin{eqnarray}
& & N_Q(i )=
  \sum_h \sum_{j=1}^{N_f} D^{(h)}  C_M(i,j)b(i){\overline
b}(j)N_Q(i)N_{\overline Q}(j) \nonumber  \\
& & +  \sum_h \sum_{j=1}^{N_f}  \sum_{k=j}^{N_f} ( 1 + \delta_{i,j} +
\delta_{i,k} ) D^{(h)} C_B(i,j,k)b(i)b(j)b(k)N_Q(i)N_Q(j)N_Q(k) ,
\nonumber \\
& & \ \  \label{S4} \\
& &N_{\overline Q}(i)=
 \sum_h \sum_{j=1}^{N_f}  D^{(h)} C_M(i,j){\overline
b}(i)b(j)N_{\overline Q}(i)N_Q(j) \nonumber \\
& & + \sum_h \sum_{j=1}^{N_f}  \sum_{k=j}^{N_f} (1+\delta_{i,j}+\delta_{i,k})
D^{(h)} C_{\overline B}(i,j,k){\overline b}(i){\overline
b}(j){\overline b}(k)N_{\overline Q}(i)N_{\overline
Q}(j)N_{\overline Q}(k) .  \nonumber \\
& & \ \  \label{S5}
\end{eqnarray}

The number of independent equations in eqs.(\ref{S4}),(\ref{S5}) is
equal to the number of independent $b(i)$ and
$\overline{b}(i)$ factors. Note, that these $2N_f$ quantities are not
adjustable parameters of the model, but they are normalization constants
determined by eqs.(4-5).

\section*{The Coalescence Factors}

 Let us consider now two particles (quarks or antiquarks) of  type
(flavor) 'a' and 'b' which form a particle (hadron) 'h'.  The rate
equation describing this  formation of particle  'h' as a  one-way
process is based  on the number  density of particles  'a' and 'b'
$(n_a, n_b)$ in the reaction volume $V$ and on
the specific cross section $\sigma$  of the $a + b  \rightarrow h$
process:
\begin{equation}
 {1 \over V} {d \over {d\tau}} \bigl( {V \,  n_h} \bigr)   =
\langle \sigma \,  v \rangle \,  n_a \,   n_b \ . \label{S6}
\end{equation}
Here  $\langle\sigma  \,   v  \rangle$  is  the  thermally averaged rate
including the  relative velocity  $v$ of  the reacting partners.
Starting such a process with  no 'h' particles at the  beginning
the  rate  of  change  is  initially constant ---
$ \langle\sigma\,  v\rangle n_a(\tau_0)n_b(\tau_0)  $  ---
thus  the  number of 'h' particles grows
linearly.  After a characteristic time $\tau$ we have
\begin{equation}
N_h   = Vn_h = { {\langle\sigma\,  v\rangle \,  \tau} \over {V} } N_a N_b
= C(a,b) N_a N_b  \label{S7}
\end{equation}
\noindent
particles. In a sudden approximation we see the effect of
different branching ratios due to competition between several
possible reaction channels producing different types of particles
(protohadrons). In this way a coalescence factor

\begin{equation}
C(a,b)  = {{\langle \sigma \,  v \rangle \,  \tau} \over {V}}  \label{S8}
\end{equation}
\vskip 0.4cm

\noindent
can be defined for each hadron. These factors influence the
relative distribution of  strange quarks between strange
mesons and singly or doubly, etc. strange baryons.

Considering the ratios of different hadron abundancies in
the final state the proper time ($\tau$) and volume ($V$) dependence
cancels, therefore the validity of the sudden approximation
can not be checked by a direct comparison with experimental
data. A long hadronization time, $\tau$, on the other hand,
would lead to chemical equilibrium between the hadrons.
In this case the relative abundances would depend only on
final state parameters and not on the branching ratios
based on specific hadronization channels. It must be
a very improbable coincidence if these two different assumptions
about the rehadronization dynamics would lead to similar
strange/non-strange hadron or meson/baryon ratios.
Thus an indirect experimental check on the sudden
approximation is provided.

We assume furthermore that the creation of a baryon from three
quarks or an antibaryon from three antiquarks is a two-step
process leading through an intermediate diquark formation in the
corresponding color triplet or antitriplet state which forms
the final baryon later
on together with a third quark or antiquark.
These diquarks must, however, be very unstable, short-lived
clusters (spatial correlations), so they may decay before forming
the baryon.  We take into account this mechanism of baryon production
by introducing the phenomenological baryon production parameter,
 $g_B$.

\newpage

\noindent
Thus the coalescence factor of a three-quark
hadron is calculated as
\begin{eqnarray}
C_B(a,b,c)  &=&  g_B
{1\over 3}  \left\{ C_M(a,b)  C_M([a+b],c)  + \right.  \nonumber \\
 & \ & \ \ \left. +   C_M(a,c)  C_M([a+c],b) + C_M(b,c) C_M([b+c],a)
 \right\},       \label{S9}
\end{eqnarray}

\noindent
where $g_B$ is assumed to be flavor blind and  we average on all
combinatoric possibility to form a baryon $N_B^{(h)}(a,b,c)$.
Here the arguments $a$, $b$ and $c$  stand for any of the flavors $u$,
$d$ or $s$ considered, respectively.

  Note that  the general scheme displayed  in
eqs.(\ref{S1} - \ref{S2}),
namely that the number of formed hadrons is proportional to  the
product of the  numbers of initials,  \hbox{$N_h=C(a,b)N_aN_b$,}
has also been conjectured  in the  framework of  a relativistic
coalescence  model  \cite{zim}.   In  this latter case,  however,  the
coalescence  factor,  $C(a,b)$,  has  been obtained in another way:
only  an  initial  and  final  state  wave function overlap in a
Gaussian  wave  packet  approximation  was  taken  into  account
instead of  using a  transition matrix  element inherent  in the
definition of the cross section involved in the ALCOR model.

There are two ingredients occurring in eq.(\ref{S8})
which have not yet been discussed: i) the reaction volume, ii) the
hadronization rate. The first is mainly restricted by the presence
of a flow preventing some candidate valence quarks 'a' and 'b'
from meeting to form a hadron 'h' if they are sitting in fluid
cells which move away from each other so fast that the local
thermal motion cannot overcome this relative velocity barrier.

In ref. \cite{SPACER} the space-time
and momentum-space evolution of the reaction at SPS
energies was investigated using the SPACER model.
It was found that the
momentum distribution of the particles can be approximated by a local
thermal distribution superimposed on a scaling flow in a finite
space-time rapidity interval.
We perform the calculation of the average reaction rate
for this case.

\section*{Hadronization Rate in the Presence of Flow}

In the presence  of a
flow we consider the relativistic J\"uttner distribution
\begin{equation}
f(x,p)  = e^{- \beta \, p \cdot u(x) }      \label{S12}
\end{equation}
  \noindent with $u_{\mu}(x)$ being  the local four-velocity  of
the flow.   Let us restrict ourselves in the following
to a scaling longitudinal flow \cite{bjorken}.
In this case the thermal averaging
of the rate can be interpreted only locally: any rate equation of
type eq.(\ref{S6}) has to be reinterpreted in terms of reacting
components from different coordinate rapidity ranges
$\eta_a$ and $\eta_b$ respectively

\begin{equation}
{1 \over V}  {d \over {d\tau}} \left( V n_h \right) = \int
d\eta_a \, d\eta_b \, \langle \sigma v \rangle_{ab} n_a n_b. \label{S14}
\end{equation}

\noindent
Clearly the total volume of the fireball increases due to a
longitudinal scaling expansion like
$ V  = \tau \pi R^2 \,  \Delta \eta$
with total coordinate rapidity extension $\Delta \eta$ and
transverse radius $R$. Assuming a space-time rapidity plateau
for a finite interval in the number
distribution of the reacting and produced components
we arrive at a modified rate equation

\begin{equation}
\tau   {d \over {d\tau} } N_h  =  {\lambda \over {\pi R^2} } N_a
N_b \ .  \label{S16}
\end{equation}

\noindent Here

\begin{equation}
 \lambda  = \int d(\eta_a - \eta_b) \, \langle \sigma v \rangle \, \Theta
\bigl( (p_a-p_b) \cdot  (x_a - x_b) \bigr)     \label{S17}
\end{equation}

\noindent is  the total  rate of reactions between all possible coordinate
rapidity cell pairs.  The constraint $(p_a - p_b) \cdot  (x_a - x_b) \ge
0$ reduces to the requirement that the relative velocity vector is
oriented opposite to the relative position vector in the center
of mass system of the colliding components 'a' and 'b'.
This constraint is necessary because  only those particles
collide which have a relative velocity  pointing towards each other.

The inclusion of this constraint after some algebraic manipulation
leads to the general expression of the averaged rate
\begin{equation}
 \lambda  = {{\beta \int d \sqrt{s} \,\, \sigma(s)
\lambda_{ab}(s)
G(\beta \sqrt{s} ) } \over { 8m_a^2m_b^2K_2(\beta m_a) K_2(\beta
m_b) } }    \label{S18}
\end{equation}

\noindent where
$ \lambda_{ab}(s)  = \left[ s-(m_a+m_b)^2 \right] \left[
s-(m_a-m_b)^2 \right]$
 and $G(\beta \sqrt{s})$ is the thermal weight
factor.  In the presence of longitudinal
scaling  flow  the  latter  can  be  written  as  an integral over
relative coordinate rapidity $\eta = (\eta_a - \eta_b)$

\begin{equation}
 G(\beta \sqrt{s})  = \int_{-\infty}^{\infty} \, d\eta \,
\sqrt{s} \, \,
{ {
K_0\left(\beta \sqrt{s}\, {\rm ch} |\eta| \right) -
K_0 \left( \beta \sqrt{s}\, {\rm ch} |\eta|  \, + \,
\beta {\lambda_{ab}^{1/2}(s) \over {2 \sqrt{s}}} {\rm sh} |\eta| \right) }
\over { \beta \lambda_{ab}^{1/2}(s) \, {\rm ch} |\eta|\, {\rm sh} |\eta| }}.
\label{S19}
\end{equation}

\vskip 0.6 truecm
Investigating the integrand of eq.(\ref{S19}) one observes a
finite width,
$\delta \eta$, in the relative coordinate rapidity $\eta$.
(This quantity, $\delta \eta$, is
equal to the relative flow rapidity because of the Bjorken scaling
assumption.) The finite width depends
on temperature $T$, on the particle
rest masses $m_a, m_b$, and on the considered energy scale
$\sqrt{s}$ in a complicated manner.

\newpage

\section*{Elementary Cross Sections}

Finally, the most  uncertain ingredient  of the  ALCOR
had\-ro\-ni\-za\-tion model, the hadronization cross section will be
discussed.  Here, until  the dynamical confinement  mechanism in
QCD has been explored, an analogy with the $p + A \rightarrow  d
+  (A-1)$  nuclear  rearrangement  (pick-up) reaction leading to
deuteron formation may give some hints \cite{schiff}.  Considering  a
final  state  in  a   Coulomb-like  potential  ---   a
simplified picture  of mesons  at intermediate
temperatures --- a fusing cross section of
\smallskip

\begin{equation}
 \sigma  = 16 m_h^2 \sqrt{\pi} \rho^3 { {\alpha^2 a } \over
{\left( 1 + (k \,  a)^2 \right)^2 }  } \label{S21}
\end{equation}
\smallskip

\noindent can  be derived  from the  above mentioned  analogy.
Here $m_h$ is the rest mass of the meson, while  $a
= 1  / (m_{ab}  \alpha) $  is the  Bohr radius  of the  bound $q
\overline{q}$ state in a $V(r) = - \alpha / r$ Coulomb potential
with $m_{ab}$ being the reduced  mass of particles 'a' and  'b'.
In our calculation we took $\alpha = 0.46$ and $T = 200$ MeV temperature
in estimating hadronization cross sections at CERN SPS  energy.
The factor $\rho  = 0.3$ fm  occurring in eq.(\ref{S21})
accounts for the medium influencing the hadron  formation and
was  taken  to  be  equal  to  the  Debye  screening  length  in
quark-gluon  plasma  at  the  above  temperature.   Finally  $k$
occurring in eq.(\ref{S21})  is the magnitude  of the relative  momentum
vector of particles 'a' and 'b' measured in their center  of
mass system $k = \lambda_{ab}^{1/2}(s) / 2\sqrt{s}$.
\bigskip

\section*{Results}
\bigskip

Well armed with the above theoretical considerations the
redistribution of different flavor quarks and antiquarks into all
possible hadrons can be calculated in the ALCOR model using
practically only three parameters: i) the total number of
quark-antiquark pairs, $N_{tot,pair} =
N_{u,pair}+ N_{d,pair} + N_{s,pair} $,
which can be determined from the measured
total charged multiplicity, ii) the parameter $g_B$ controlling
the baryon formation and iii) the strangeness
production factor $g_s$.
The dependence of the final results on the other parameters
(e.g. $T$, $\alpha$) within their physically acceptable interval is small.
 All further
results, such as the number of hyperons, kaons, etc. are predictions
of the ALCOR model.

\newpage
\ \

\bigskip

\begin{center}
\begin{tabular}{||c||c||c||c|c|c||}
\hline {\bf S+S} &
 {\bf DATA}  & {\bf ALCOR}& {\bf HIJ.01} &{\bf RQMD } & {\bf QGSM} \\
\hline
\hline
 $h^{-}$  & $98. \pm 3$ & 100.2  & 88.80 & 110.2 & 120. \\
\hline
\hline
 $\pi^+$    &  &88.14  &  & & \\
\hline
 $\pi^0$    &  &88.14  &  & & \\
\hline
 $\pi^-$    & $91. \pm 3$  &88.14  &79.60  & & \\
\hline
 $K^+$    & $12.5 \pm 0.4 $  &12.70  &8.43  & & \\
\hline
 $K^0$    &  &12.70  &  & & \\
\hline
 ${\overline K}^0$    &  &\ 6.36  &  & & \\
\hline
 $K^-$    & $6.9 \pm 0.4 $  &\ 6.36  &6.27  & & \\
\hline
 $K^0_{S}$   &$ 10.50 \pm 1.7$  &\ 9.53  &\ 7.23  &10.0 & 7.4 \\
\hline
\hline
 $p^+$    &  &22.04  &  & & \\
\hline
 $n^0$    &  &22.04  &  & & \\
\hline
 $\Sigma^+$    &  &\ 1.71  & &  &  \\
\hline
 $\Sigma^0$    &  &\ 1.71  & &  &  \\
\hline
 $\Sigma^-$    &  &\ 1.71  &  & &  \\
\hline
 $\Lambda^0$   &  &\ 8.54  &\ 4.58  & 7.76 &  4.7 \\
\hline
 $Y^0 = \Sigma^0 + \Lambda^0$    &$ 9.4 \pm 1.0$  &10.25  & &  &  \\
\hline
 $\Xi^0$   &  &\ 1.13  &  & &  \\
\hline
 $\Xi^-$   &  &\ 1.13  &\ 0.04 & &  \\
\hline
 $\Omega^{-}$   &  &\ 0.19  &  & &  \\
\hline
\hline
 ${\overline p}^-$   &  &\ 2.35  &  & &  \\
\hline
 ${\overline n}^0$   &  &\ 2.35  &  & &  \\
\hline
 ${\overline \Sigma}^-$   &  &\ 0.40  &  & &  \\
\hline
 ${\overline \Sigma}^0$   &  &\ 0.40  &  & &  \\
\hline
 ${\overline \Sigma}^+$   &  &\ 0.40  &  & &  \\
\hline
 ${\overline \Lambda}^0$   & &\ 1.98  &\ 0.86& & 0.35  \\
\hline
 ${\overline Y}^0 = {\overline \Sigma}^0 + {\overline \Lambda}^0$

 &$2.20\pm 0.4$  &\ 2.38  &  & &  \\
\hline
 ${\overline \Xi}^0$   &  &\ 0.57  &  & &  \\
\hline
 ${\overline \Xi}^+$   &   &\ 0.57  &\ 0.06  & &  \\
\hline
 ${\overline \Omega}^{+}$   &  &\ 0.21  &  & &  \\
\hline
\hline
\end{tabular}
\end{center}

\vskip 0.8cm
\noindent {\bf Table 1:}
Hadron multiplicities for $S+S$ collision observed experimentally
and obtained in the ALCOR and in the HIJING, RQMD, QGSM models
(early non-collective string model versions, taken from refs.
\cite{HIJ01} \cite{QGSM}) at 200 GeV/nucleon bombarding energy.

\vfill
\newpage

 Table  1  shows  results  for the S+S reaction at 200  GeV/nucleon bombarding
energy  together  with the  experimental data.   Column 1
displays  multiplicities  observed   in experiment NA35
\cite{NA3593}, \cite{NA3594}.
Column 2  contains  the  hadron   multiplicities
predicted by the ALCOR model.
Adjusting the three above parameters of ALCOR one obtains fairly good
agreement with the experimental data for particles
$\pi^-,  K^+,  K^-,   K^0_s,  Y^0  $ and
${\overline Y}^0 $.
(Here we introduced the notation $Y^0 = \Lambda^0 + \Sigma^0$ and
${\overline Y}^0 = {\overline \Lambda}^0 + {\overline \Sigma}^0$ for the
measured neutral strange and anti-strange baryons .)
We took into account that
the number of participant nucleons was measured to be
$N_{partic}^{SS} = 51$ \cite{particip}.  We used the parameters
$N_{tot,pair}^{SS}=158.1$ with $g_s=0.255$ for the strangeness production,
and $g_B=0.04$ for the baryon formation .
For comparison we display some results of the HIJING, RQMD and
QGSM models in Columns 3, 4 and 5 taken from refs.
\cite{HIJ01} \cite{QGSM} (early non-collective string model versions).
Table 2 shows  results for the strange baryon and anti-baryon ratios.
Column 1 displays the experimental data \cite{WA94} and
Column 2 contains the ALCOR predictions. The agreement
seems to be good.

\vskip 0.8cm

\begin{center}
\begin{tabular}{||c||c|c||}
\hline {\bf S+S} &  {\bf WA94}  & {\bf ALCOR}  \\
\hline
\hline
 ${\overline {Y^0}}/Y^0$   & $0.23 \pm 0.01$ & 0.23   \\
\hline
 ${\overline {\Xi^-}}/\Xi^-$& $0.55 \pm 0.07$ & 0.50   \\
\hline
 $\Xi^-/Y^0$               & $0.09 \pm 0.01$ & 0.11   \\
\hline
 ${\overline {\Xi^-}}/{\overline {Y^0}}$ &$0.21 \pm 0.02$ & 0.24\\
\hline
\hline
\end{tabular}
\end{center}
\vskip 0.8cm

\noindent {\bf Table 2:}
Strange baryon and anti-baryon ratios measured by
WA94 Collaboration \cite{WA94}
and obtained from ALCOR for $S+S$ collision at 200 GeV/nucl bombarding energy.

\vskip 0.5cm

{}From this experience we conjecture that the above rehadronization
process is {\bf locally quick}: the quarks and antiquarks (including the
fragmented   gluons)  appear   in   hadrons   according  to simple
kinematic rules and production branching ratios.
This finding, however, does not settle the question whether
quark  matter   with  any  degree   of  collectivity has been formed
or other hadronization  processes  including  strings
or color ropes \cite{rope} are  the source of this amount of quarks.
Local thermal distribution  of the  different  quark flavors  and the
presence of a  longitudinal flow cannot, on the other hand, be excluded
on the basis of this experimental data.

\newpage

To obtain results for other reactions, we assume a scaling for the produced
quark-antiquark pairs as
\begin{equation}
N_{tot,pair}^{AA} (\sqrt{s}) =
\left( { N_{partic}^{AA} \over N_{partic}^{SS}}\right)^\alpha
N_{tot,pair}^{SS}, (\sqrt{s}) \label{S22}
\end{equation}
\noindent where $N_{partic}^{AA}$ are the number of participant nucleons in the
A+A collision.

The scaling exponent $\alpha$ may have the value $\alpha=1$,
or, for more collective production processes one expects $\alpha = 4/3$.
Furthermore we shall use the  values for $g_B$ and $g_s$ obtained above.
If one of these parameters has to be changed
in order to  approve the correspondence of the calculated particle numbers
with the experimental ones,  than this change
must have a physical interpretation.

\ \

\vskip 0.8cm

\begin{center}
\begin{tabular}{||c||c|c||c|c||}
\hline & {\bf p+n}  & {\bf ALCOR}& {\bf p+p} &{\bf ALCOR }  \\
\hline
\hline
 $h^{-}$  & $3.23 \pm 0.02$ & 3.24  & $2.85 \pm 0.03$ & 2.87 \\
\hline
\hline
 $\pi^+$  & $3.01 \pm 0.04$ & 3.03  & $3.22 \pm 0.12$ & 3.43 \\
\hline
 $\pi^0$  & $3.06 \pm 0.25$ & 3.03  & $3.34 \pm 0.24$ & 3.01 \\
\hline
 $\pi^-$  & $3.01 \pm 0.04$ & 3.03  & $2.62 \pm 0.06$ & 2.67 \\
\hline
 $K^+$    & [0.24]          & 0.28  & $0.28 \pm 0.06$ & 0.28 \\
\hline
 $K^-$    & [0.17]          & 0.12  & $0.18 \pm 0.05$ & 0.12 \\
\hline
$K^0_{S}$ & [0.20]          & 0.20  & $0.17 \pm 0.01$ & 0.20 \\
\hline
\hline
 $p^+$    & $1.00 \pm 0.08$ & 0.88  & $1.34 \pm 0.15$ & 1.10 \\
\hline
 $n^0$    & $1.00 \pm 0.08$ & 0.88  & $0.61 \pm 0.30$ & 0.65 \\
\hline
 $Y^0=\Sigma^0 + \Lambda^0$

          & [0.096]         & 0.23  & $0.096 \pm 0.01$& 0.22 \\
\hline
\hline
 ${\overline p}^-$

          & (0.05)          & 0.03  & $0.05 \pm 0.02$ & 0.03 \\
\hline
 ${\overline n}^0$

          & (0.05)          & 0.03  & (0.05)          & 0.03 \\
\hline
 ${\overline Y}^0 = {\overline \Sigma}^0 + {\overline \Lambda}^0$

          & (0.013)         & 0.019 & $0.013\pm 0.004$& 0.019\\
\hline
\hline
\end{tabular}
\end{center}
\vskip 0.8cm

\noindent {\bf Table 3:}
Hadron multiplicities for $p+n$ and $p+p$
collision observed experimentally
\cite {Gazd} and obtained in the ALCOR  models
at 200 GeV/nucleon bombarding energy. The numbers in square brackets
are the average of the corresponding $p+p$ and $n+n$ results.
The numbers in normal brackets are assumed values with $50\%$ uncertainty
\cite{Gazd}.

\newpage

With these assumption we calculated the particle production also for
nuc\-leon-\-nucleon collisions namely for proton-proton and proton-neutron
collisions at 200 GeV/nucleon energy. For this case, in order to get
the number of strange hadrons
correctly, we had to decrease the strangeness production factor. Table 3.
shows the ALCOR results for $p+n$ and $p+p$ collisions
together with  the experimental values from
Ref. \cite{Gazd}. In both cases we used the same parametrization,
namely $N_{partic}^{NN}=2$,  $N_{tot,pair}^{NN}=4.5$
with a decreased strangeness formation factor $g_s=0.16$
and an unchanged baryon formation factor $g_B=0.04$. The scaling
exponent is $\alpha=1.1$. (In the two cases only the
flavour component of the incoming participant
nucleons were different, i.e. $p+p$ and $p+n$, respectively.)
One can observe, that the calculated meson and non-strange baryon
numbers are not far from the measured ones. To achive this we had to
decrease the strangeness production factor, $g_s$, from its value in $S+S$
reaction. Thus we may conclude, that {\bf in the $S+S$ reaction the strangeness
production is enhanced with respect to the nucleon-nucleon reaction
by a factor of 1.6}.
Furthermore we may observe from Table 3., that the calculated
$Y^0=\Sigma^0+\Lambda^0$ and
${\overline Y}^0={\overline \Sigma}^0+{\overline \Lambda}^0$
numbers deviate  essentially from the measured ones.
Namely ALCOR overpredicts the production of these particles. This
discrepancy can be interpreted by assuming that
{\bf the hadronization mechanism
in the $S+S$ reaction is different from that for the nucleon+nucleon reaction}
and this can be seen undoubtedly on strange baryon production.
In fact, the ALCOR model assumes, that there may be quark exchanges between
neighbouring excited objects (strongly packed strings, etc.) allowing larger
role to the combinatoric possibilities.
Thus one may describe this  case as a formation of
a {\bf semi-deconfined state} of the matter.
However, in nucleon-nucleon collision
this effect is missing.

Using  the parametrization of S+S collision and the above scaling law,
we may make
predictions  for  the  Pb+Pb  collision  at 160 GeV/nucleon energy.
In this case the number of participant nucleons is
$N_{part}^{PbPb} = 390 \pm 10$
obtained from Monte-Carlo simulations \cite{MarekP}.
We will use the mean value.
During the extrapolation of
the  total  number  of  produced quark-antiquark pairs from the S+S collision
one needs to consider an energy rescaling. We will assume a logaritmic one:
\begin{equation}
{{N_{tot,pair}^{AA} (\sqrt{s_1}) } \over
 {N_{tot,pair}^{AA} (\sqrt{s_2}) }}  =
{ {\ln{\sqrt{s_1}}} \over {\ln{\sqrt{s_2}}} }
\end{equation}
This rescaling yields a $\approx 4 \%$ correction and
one obtains $N_{tot,pair}^{PbPb}=1164$ in the linear case ($\alpha=1$)
and $N_{tot,pair}^{PbPb}=2294$ in the more collective one ($\alpha=4/3$).
Table 4 displays the predictions of ALCOR  for both scaling cases
and the last column contains some results from HIJING taken from ref.
\cite{HIJ01}.

\newpage
\ \

\bigskip

\begin{center}
\begin{tabular}{||c||c|c|c||}    \hline
\hline {\bf Pb+Pb} &
 {\bf ALCOR} ($\alpha=1$)& {\bf ALCOR} ($\alpha=4/3$)&{\bf HIJ.01} \\
\hline
\hline
 $h^{-}$  & 730.41  & 1231.0  &725.15 \\
\hline
\hline
 $\pi^+$  & 603.87 &  973.38 & \\
\hline
 $\pi^0$  & 618.95 &  988.84 & \\
\hline
 $\pi^-$  & 634.68 & 1004.7 &621.75   \\
\hline
 $K^+$  & \ 84.15 & 133.35 & \\
\hline
 $K^0$  & \ 84.15 & 133.35 & \\
\hline
 ${\overline K}^0$  &\ 41.65  &\ 87.96  &  \\
\hline
 $K^-$  &\ 41.65  &\ 87.96  & \\
\hline
 $K^0_{S}$  & \ 62.90  &110.66  &\ 54.86   \\
\hline
\hline
 $p^+$  &170.90  &218.97  & \\
\hline
 $n^0$  &188.57  &234.84  & \\
\hline
 $\Sigma^+$  &\ 12.86  &\ 22.31  &  \\
\hline
 $\Sigma^0$  &\ 13.63  &\ 23.25  & \\
\hline
 $\Sigma^-$  &\ 14.43  &\ 24.22  & \\
\hline
 $\Lambda^0$  &\ 68.20  &116.31 &\ 36.44  \\
\hline
 $\Xi^0$  &\ \ 8.82  &\ 20.35  & \\
\hline
 $\Xi^-$  &\  \ 8.89 &\ 20.45  &\ \ 0.22   \\
\hline
 $\Omega^{-}$  &\ \ 1.48  &\ \ 4.60  &  \\
\hline
\hline
 ${\overline p}^-$  &\ 25.07 &\ 76.41  & \\
\hline
 ${\overline n}^0$  &\ 25.07 &\ 76.41  & \\
\hline
 ${\overline \Sigma}^-$  &\ \ 4.18  &\ 12.58  & \\
\hline
 ${\overline \Sigma}^0$  &\ \ 4.18  &\ 12.58  & \\
\hline
 ${\overline \Sigma}^+$  &\ \ 4.18  &\ 12.58  & \\
\hline
 ${\overline \Lambda}^0$  &\ 20.93  &\ 62.90  &\ \ 5.93   \\
\hline
 ${\overline \Xi}^0$  &\ \ 5.98  &\ 17.71  &   \\
\hline
 ${\overline \Xi}^+$  &\ \ 5.98  &\ 17.71  &\ \ 0.40 \\
\hline
 ${\overline \Omega}^{+}$&\ \ 2.20&\ \ 6.42&  \\
\hline
\hline
\end{tabular}
\end{center}

\vskip 0.8cm

\noindent {\bf Table 4:}
Hadron multiplicities predicted by the ALCOR model assuming different
scaling laws discussed in the text. For comparison,
predictions of the HIJING model taken from ref. \cite{HIJ01} are also
selected.

\newpage

\section*{Conclusion}

\bigskip

We have applied  the hadronization model, ALCOR, for the analysis of the
experimental data of $p+p$, $p+n$ and $S+S$  collisions
and for the prediction of hadron multiplicities in
the $Pb+Pb$ collision at CERN SPS energy.

This model uses dynamical coalescence
probabilities obtained from hadro\-pro\-duction
reaction rate factors obtained on a one dimensional scaling flow
background. The model is sensitive to only three free parameters:
the total number of quark-antiquark pairs produced,
an in-medium baryon formation factor $g_B$, and
the strangeness production factor $g_s$,
fitted to the experimental results of the
reaction S+S. (The effect of changes in common factors are removed by the
normalization constants.)
With no further adjustment we have been able to
describe several observed hadron multiplicities in agreement
with the  NA35 and WA94 experiments. We could clearly identify the
strangeness enhancement in $S+S$ collision with respect to the nucleon-nucleon
reaction.
\smallskip

Furthermore, for $p+p$ and $p+n$ collisions the ALCOR model could not
reproduce the measured strange baryon and antibaryon numbers on the basis of
$S+S$ collision. We interpret this discrepancy as a signature
showing that the hadronization mechanism
in the $S+S$ reaction differs from that for the nucleon+nucleon reaction.
We infer from this comparison that in $S+S$ collision a sort of
semi-deconfined state
of the matter was formed which can be characterized by enhanced quark
exchanges between neighbouring excited objects. This is a basic assumption
in the ALCOR model.

\medskip

We made detailed predictions for hadron multiplicities in a Pb+Pb
reaction assuming two different scaling exponent. There is an agreement
between the HIJING and the ALCOR model (in linear case, $\alpha=1$)
for prediction of total negative particle, pion and kaon
multiplicities. However, the strange baryon and anti-baryon yields
differ from the HIJING result. Now we look forward to compare these
predictions with the results of the NA49 CERN SPS experiment
to determine the details of hadronization processes.
Possible agreement would support our
statement about the formation of semi-deconfined state of the matter
in high energy heavy ion collisions.

\vskip 1cm

\section*{ Acknowledgement}

Discussions with T. Cs\"org\H o, M. Gazdzicki,
K. Kadija, H. Sorge,
H. Str\"obele and X. N. Wang are acknowledged.
This work was supported by the National Scientific
Research Fund (Hungary), OTKA
No.T014213 and by the U.S. - Hungarian Science and Technology
Joint Fund, No. 378/93.


\newpage

\small

\begin{thebibliography}{99}


\bibitem{Wang} X.N.Wang, M.Gyulassy, Phys. Rev. {\bf D44} 3501 (1991)


\bibitem{parton}  K.Geiger, B.M\"uller, Nucl. Phys. {\bf B369} 600 (1992)


\bibitem{qchem} T.S.Bir\'o, J.Zim\'anyi, Phys. Lett. {\bf B113} 6 (1982)


\bibitem{hchem} I.Montvay, J.Zim\'anyi, Nucl. Phys. {\bf A316} 490 (1979)


\bibitem{Werner} K.Werner, Phys. Reports {\bf 232} 87 (1993)


\bibitem{rope} T.S. Bir\'o, J. Knoll, H.B. Nielsen, Nucl.
Phys. {\bf B245} 449 (1984)

\bibitem{ALCOR} T.S. Bir\'o, P. L\'evai, J. Zim\'anyi, Phys. Lett.
{\bf B347} 6 (1995)

\bibitem{recom} T.S.Bir\'o, J.Zim\'anyi, Nucl. Phys. {\bf A395} 525 (1983)


\bibitem{gfrag} P.Koch, B.M\"uller, J.Rafelski, Phys. Report
 {\bf 142} 167 (1986)


\bibitem{zim} J.Zim\'anyi, P.L\'evai, W.Greiner, U.
Heinz, in preparation


\bibitem{SPACER} T.Cs\"org\H{o}, J.Zim\'anyi, J.Bondorf, H.Heiselberg,
Phys. Lett. {\bf B222} 115 (1989)


\bibitem{bjorken} J.D.Bjorken, Phys. Rev. {\bf D27} 140 (1982)


\bibitem{schiff} e.g. L.I.Schiff, Quantum Mechanics,Chapt. 34,
 McGraw-Hill, New York, 1955

\bibitem{HIJ01} V. Topor Pop at al., WA94 Coll.,
DFPD-94-NP-49 Preprint (hep-ph/9407262)


\bibitem{QGSM} N.S. Amelin, L.V. Bravina, L.P. Csernai, V.D.
Toneev, K.K. Gudima, S.Yu. Sivoklokov, Phys. Rev {\bf C47}
2299 (1993)

\bibitem{NA3593} J.Baechler et. al.  Z.Phys. {\bf C58} 367 (1993)


\bibitem{NA3594} NA35 Coll., T. Alber et al., IKF-HENPG/1-94  Univ. Frankfurt
Preprint

\bibitem{particip} J. B\"achler et.al., NA35 Coll., Phys. Rev. Lett.
{\bf 72} 1419 (1994)

\bibitem{WA94} WA94 Collaboration, data presented in
S'95 by O. Villalobos-Baillie.

\bibitem{Gazd} M. Gazdzicki, O. Hansen, Nucl. Phys. {\bf A528} 754 (1991)

\bibitem{MarekP} M. Gazdzicki, private communication.

\end{thebibliography}
\end{document}